\documentclass{emulateapj}

\usepackage{amsmath}
\usepackage{amssymb}
\usepackage{graphicx}

\newcommand{\kepler}{\texttt{Kepler}}
\newcommand{\kms}{\ensuremath{\mathrm{km~s}^{-1}}}

\newcommand{\Nifs}{\ensuremath{^{56}\mathrm{Ni}}}

\newcommand{\Msun}{{\ensuremath{\mathrm{M}_{\odot}}}}

\newcommand{\vs}{\ensuremath{v_{\rm s}}}
\newcommand{\ergss}{\ensuremath{{\rm ergs~s}^{-1}}}
\newcommand{\gcc}{\ensuremath{{\rm g~cm}^{-3}}}

\shortauthors{Kasen, Woosley, \& Heger}
\shorttitle{Pair Instability Supernovae}

\begin{document}

\title{Pair Instability Supernovae:  Light Curves, Spectra, and Shock Breakout}
\author{Daniel Kasen\altaffilmark{1,2} S.E. Woosley\altaffilmark{3}
and A. Heger\altaffilmark{4}}

\altaffiltext{1}{Departments of Physics and Astronomy, University of California, Berkeley; \email{kasen@berkeley.edu}}
\altaffiltext{2}{Nuclear Science Division, Lawrence Berkeley National Laboratory}
\altaffiltext{3}{Department of Astronomy and Astrophysics, University of California, Santa Cruz}
\altaffiltext{4}{University of Minnesota}

\begin{abstract}
 For the initial mass range ($140 < M < 260~\Msun$) stars die in a
  thermonuclear runaway triggered by the pair-production instability.
  The supernovae they make can be remarkably energetic (up to $\sim
  10^{53}$ ergs) and synthesize considerable amounts of radioactive
  isotopes.  Here we model the
  evolution, explosion, and observational signatures of representative
  pair-instability supernovae (PI~SNe) spanning a range of initial
  masses and envelope structures. The predicted light curves last for
  hundreds of days and range in luminosity, from very dim to extremely bright $L \sim 10^{44}$
  ~ergs/sec.  The most
  massive events are bright enough to be seen at high redshift, but
  the extended light curve duration ($\sim 1$~year) -- prolonged by
  cosmological time-dilation -- may make it difficult to detect them
  as transients.  An alternative approach may be to search for the
  brief and luminous outbreak occurring when the explosion shock wave
  reaches the stellar surface. Using a multi-wavelength
  radiation-hydrodynamics code we calculate that, in the rest-frame,
  the shock breakout transients of PI~SNe reach luminosities of
  $10^{45}-10^{46}$~\ergss, peak at wavelengths $\sim 30-170$~\AA, and
  last for several hours.   We explore the detectability of PI SNe
  emission at high redshift, and discuss how observations
   of the light curves, spectra, and
  breakout emission can be used to constrain the mass, radius, and
  metallicity of the progenitor.
  \end{abstract}

\section{Introduction}

\begin{deluxetable*}{lrrrccrrrrc} 
\tablewidth{0pt}
\tablecaption{Pair Instability Supernova Explosion Models}
\tablehead
{
\colhead{Name}                & 
\colhead{$M_i$\tablenotemark{a}}               & 
\colhead{$M_f$}               & 
\colhead{$M_{\rm He}$}        & 
\colhead{$\rho_{\rm b}$\tablenotemark{b}}  &
\colhead{$T_{\rm peak} (10^9$K)}      & 
\colhead{$M_{\rm ej}$}        & 
\colhead{energy (B)}         & 
\colhead{$M_{\rm ni}$}        &
\colhead{$R_0 (10^{12}$cm)}              &
\colhead{L\tablenotemark{c}}                 
}
\startdata                                                
B150  &  150 &   150  &   66.9  & 1.87   &   3.51  &   68.9  &  5.85   &   0.00  &   4.63    &   1.50  \\
B175  &  175 &   175  &   84.3  & 2.31   &   3.88  &   90.9  &  14.6   &   0.00  &   6.24    &   1.82  \\
B200  &  200 &   200  &   96.9  & 2.86   &   4.29  &   200   &  27.8   &   1.90  &   6.57    &   2.13  \\
B225  &  225 &   225  &  110.1  & 3.75   &   4.75  &   225   &  42.5   &   8.73  &   9.79    &   2.46  \\
B250  &  250 &   250  &  123.5  & 5.59   &   5.38  &   250   &  63.2   &  23.10  &   13.1    &   2.79  \\
R150  &  150 & 142.9  &   72.0  & 2.16   &   3.70  &  142.9  &   9.0   &   0.07  &    162    &   1.42  \\
R175  &  175 & 163.8  &   84.4  & 2.66   &   4.10  &  163.8  &  21.3   &   0.70  &    174    &   1.69  \\
R200  &  200 & 181.1  &   96.7  & 3.32   &   4.56  &  181.1  &  33.0   &   5.09  &    184    &   1.76  \\
R225  &  225 & 200.3  &  103.5  & 4.88   &   5.15  &  200.3  &  46.7   &   16.5  &    333    &   2.10  \\
R250  &  250 & 236.3  &  124.0  & 9.45   &   6.16  &  236.3  &  69.2   &   37.86 &    225    &   2.60  \\
He070 &   70  & 70.0     &   70.0  & 2.00   &   3.57  &   70.0  &   8.2   &    0.02 &      -    &      -  \\
He080 &   80  & 80.0     &   80.0  & 2.32   &   3.88  &   80.0  &  17.5   &    0.19 &      -    &      -  \\
He090 &   90  & 90.0     &   90.0  & 2.70   &   4.20  &   90.0  &  28.6   &    1.15 &      -    &      -  \\
He100 &  100  & 100.0    &  100.0  & 3.20   &   4.53  &  100.0  &  40.9   &    5.00 &      -    &      -  \\
He100F & 100  & 100.0    &  100.0  & 2.98   &   4.44  &  100.0  &  40.2   &    3.64 &      -    &      -  \\           
He110 &  110  & 110.0    &  110.0  & 4.08   &   4.93  &  110.0  &  55.6   &   12.12 &      -    &      -  \\
He120 &  120  & 120.0    &  120.0  & 5.42   &   5.39  &  120.0  &  70.6   &   23.83 &      -    &      -  \\  
He130 &  130  & 130.0    &  130.0  & 9.01   &   6.17  &  130.0  &  86.7   &   40.32 &      -    &      -  \\
\enddata
\tablenotetext{a}{this and all other masses in units of \Msun}
\tablenotetext{b}{bounce density in $10^{-6}$~\gcc}
\tablenotetext{b}{presupernova luminosity in  $10^{40}$~\ergss}
\label{tab:models}
\end{deluxetable*}

Models of metal-free star formation suggest that the first stars to
form in the universe were likely quite massive, $M > 100$~\Msun
\citep{Bromm_1999,Abel_2000,Nakamura_2001}.  The low metallicity
of these star may have allowed them to retain much of their mass
throughout their evolution.  The most massive objects ($M >
260~\Msun$) are thought to end their lives by direct collapse to a
black hole, with no associated supernova \citep{Heger_Death}.  But
stars with initial masses between $\sim 140$ and $260$~\Msun\ fall
prey to the pair-production instability and explode completely
\citep{Barkat_PPSN,Rakavy_1967,Bond_1984,Umeda_2002,Heger_PPSN,
Waldman_2008,Moriya_2010, Fryer_2010}.  The high core temperatures lead to
the production of e$^+$/e$^-$ pairs, softening the equation of state
and leading to collapse and the ignition of explosive oxygen burning.
The subsequent thermonuclear runaway reverses the collapse and ejects
the entire star, leaving no remnant behind.  The explosion physics is
fairly well understood and can be modeled with fewer uncertainties
than for other supernova types.

The predicted explosion energy of pair instability supernovae (PI~SNe)
is impressive, nearly $10^{53}$~ergs for the most massive stars
\citep{Heger_PPSN}. Radioactive \Nifs\ can be synthesized in
abundance, up to 40~\Msun\ of it.  This is almost 100 times the energy
and \Nifs\ yield of a typical Type~Ia supernova (SN~Ia).  The light
curves of the most massive PI~SNe are then expected to be very luminous ($\sim
10^{43}-10^{44}$~ergs/sec) and long-lasting ($\sim 300$~days).  Deep
searches could potentially detect these events in the early universe,
offering a means of probing the earliest generation of stars
\citep{scano_PPSN}.  

Interest in PI~SNe has recently been renewed by the discovery, in the
more nearby universe, of several supernovae of extraordinary
brightness \citep{Knop_99as,Quimby_05ap, barbary_09, Quimby_09,Smith_06gy, 
Gezari_2009}.  In some cases, the high luminosity can be attributed
to an interaction of the supernova ejecta with a surrounding
circumstellar medium \citep{Smith_McCray}.  But other events show no
clear signatures of interaction, and appear to have synthesized large
quantities of \Nifs.  The speculation is that some of these events
represent the pair instability explosion of very massive stars,
perhaps having formed in pockets of relatively lowly enriched gas.
So far, the most promising candidate appears to be SN~2007bi
 \citep{gal_yam_2009,Young_09}, a Type~Ic supernova that
was both over-luminous and of extended duration.

In this paper, we model the stellar evolution (\S\ref{sec:evolve}) and
explosion (\S\ref{sec:explode}) of PI~SN models spanning a range of
initial masses and envelope structures.  Using a
radiation-hydrodynamics code, we then calculate the very luminous
breakout emission that occurs when the explosion shock wave first
reaches the surface of the hydrogen envelope ({\S\ref{sec:bo}}).  We 
follow with time-dependent radiation transport calculations of the
broadband light curves (\S\ref{sec:lc}) and spectral time-series
(\S\ref{sec:spectra}).  The models illustrate how the observable
properties of PI~SNe can be used to constrain the mass, radius, and
metallicity of their progenitors stars.  They also allow us to
evaluate the prospects of discovering these events in upcoming
observational surveys (\S\ref{sec:detect}).

The detectability of PI~SNe was explored previously by
\cite{scano_PPSN}, who used a grey flux-limited diffusion method to
model the light curves.  In this paper, we have generated a new set of
more finely resolved models which explores the parameter space in a
systematic way. We have also significantly improved the radiative
transfer calculations by using a multi-wavelength implicit Monte Carlo
code which includes detailed line opacities.  This allows us to 
generate synthetic spectra and to predict the color
evolution and K-correction effects.
While \cite{scano_PPSN} focused on the long duration
light curves, we model here as well the brief and
luminous transient at shock breakout, and consider whether that might
be a useful signature for finding PI~SNe soon after they explode.

\section{Evolution and Explosion}

\begin{figure*}
\includegraphics[width=6.in]{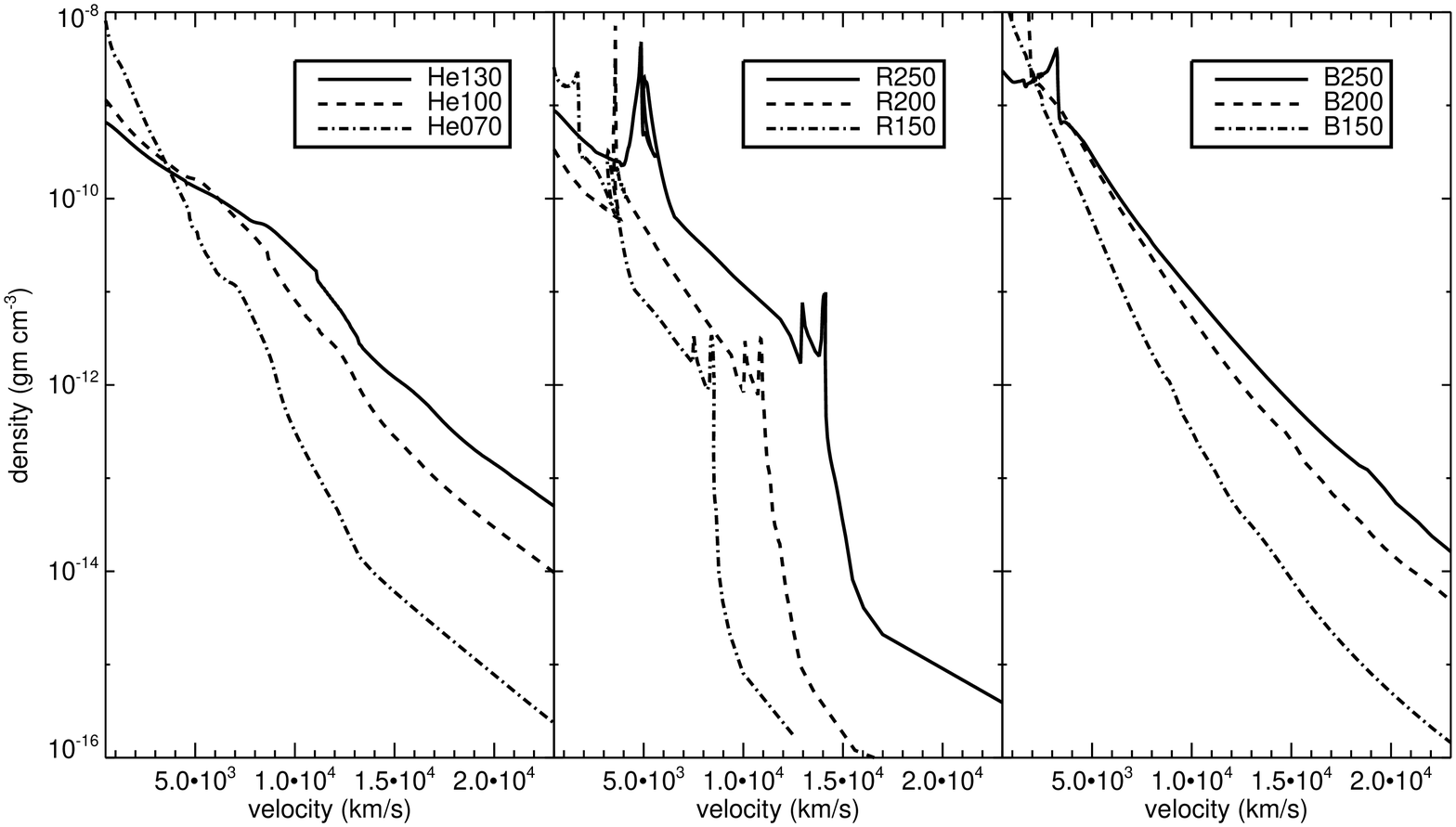}
\caption{Ejecta density profiles of  PI~SN models 10 days after explosion, when the 
expansion is nearly homologous ($v \propto r$).
\label{Fig:density} }
\end{figure*}

\begin{figure*}
\begin{center}
\includegraphics[width=2.2in]{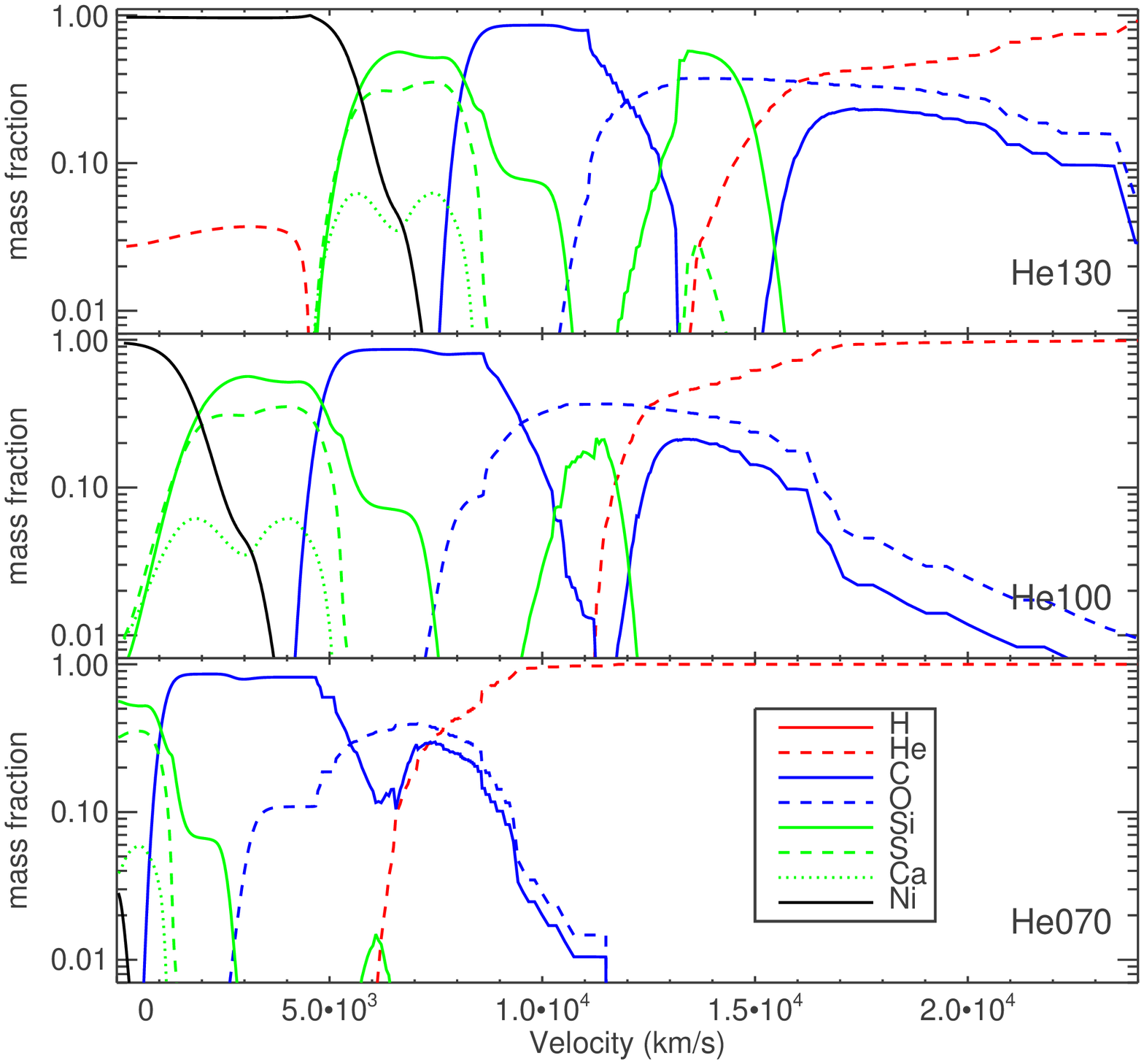}
\includegraphics[width=2.2in]{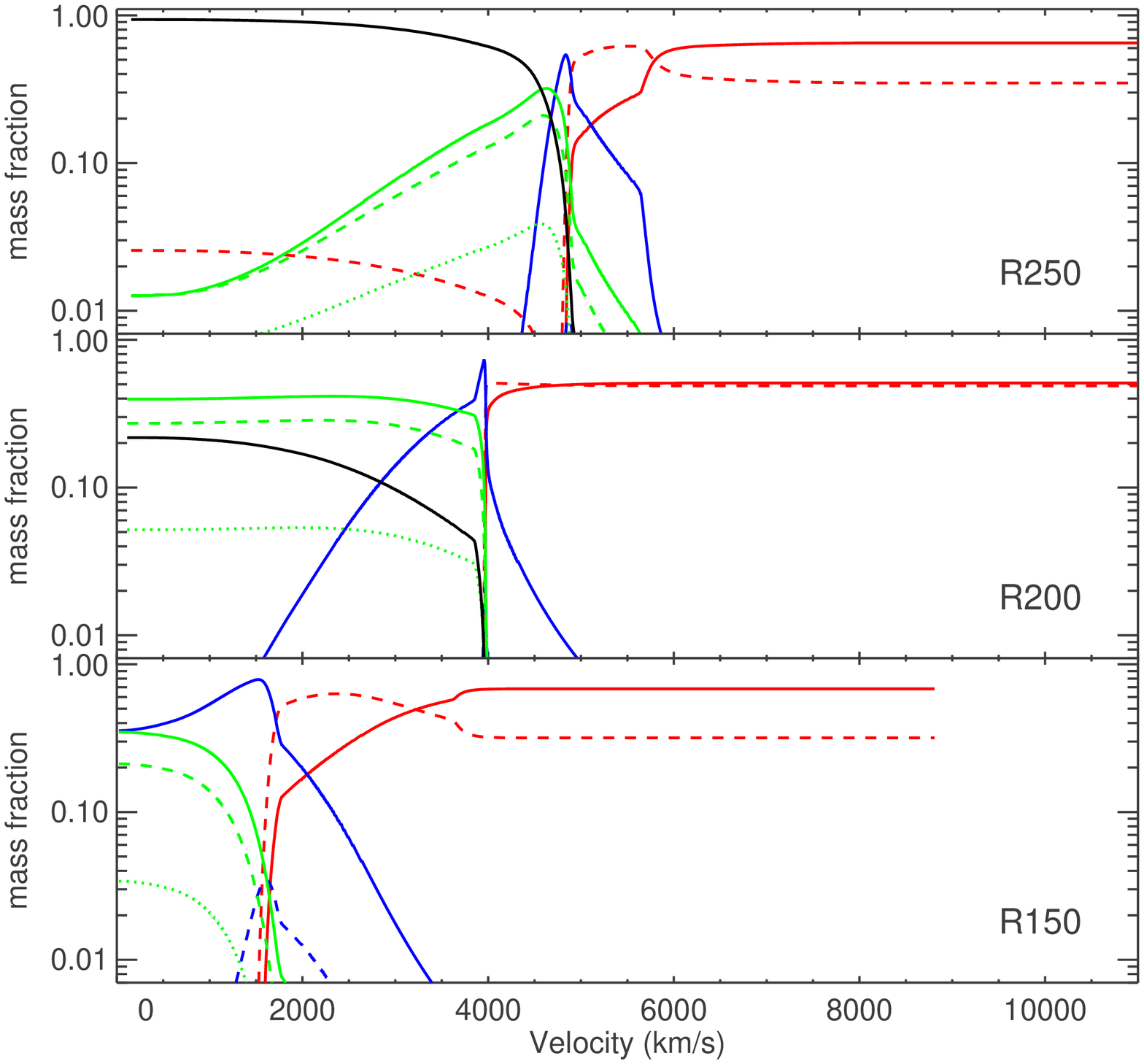}
\includegraphics[width=2.2in]{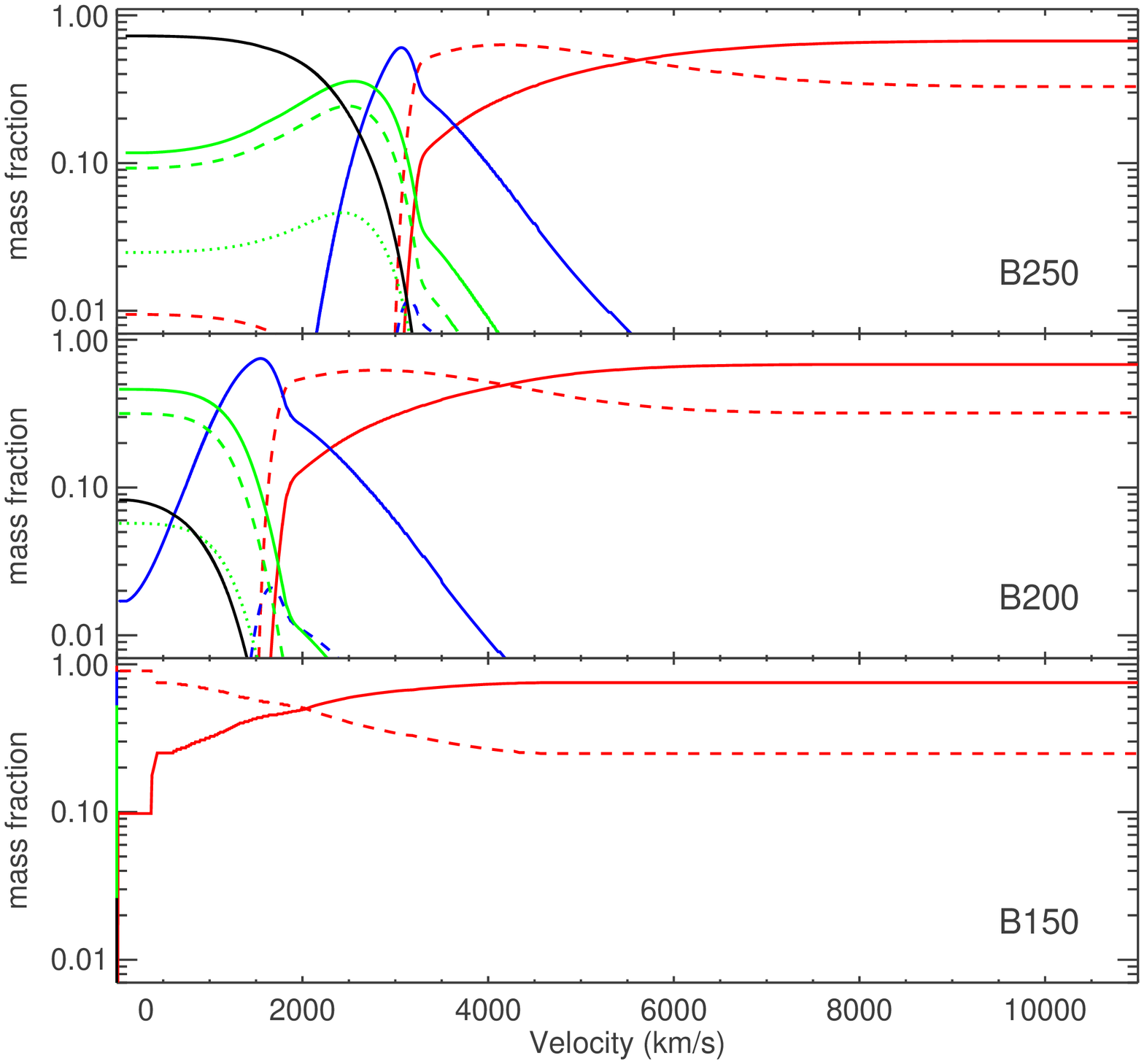}
\end{center}
\caption{Ejecta abundance structures of three of the PI~SN models     
in the homologous expansion phase.  An artificial smearing of the
compositional interfaces has been applied to mimic multi-dimensional
mixing processes.
\label{Fig:abun} }
\end{figure*}

\subsection{Stellar Evolution Models}
\label{sec:evolve}

The electron-positron pair-instability supernova mechanism was
originally discussed by \citet{Rakavy_PP} and \citet{Barkat_PPSN}, and
has since been explored with a number of numerical and analytic models
\citep[see][and references therein]{Heger_PPSN}.  Most recently,
\citet{Heger_PPSN} [hereafter HW02] explored the explosion of
non-rotating bare helium cores with a range of masses from 64 to 133
\Msun.  A subset of these models, with masses 70 - 130 \Msun\ (in
steps of 10 \Msun) will be examined here.

Models of bare helium cores, though computationally expedient, likely
do not fully represent the class of PI~SNe, as not all stars will have
lost their hydrogen envelopes to mass loss or binary mass exchange
just prior to exploding.  We therefore evolved a new set of models
especially for this study of the breakout transients, light curves,
and spectra.\footnote{These models are available to others seeking to
  carry out similar studies.} They consist of five models each of
``hydrogenic" (i.e., possessing hydrogen) stars in the main sequence mass range 150 to 250 \Msun,
with initial metallicities of 0 and 10$^{-4}$ times solar.  The
surface zoning of these models was chosen much finer than those of
HW02 in order to facilitate calculation of the shock breakout
emission.  In addition, one helium core of mass 100 \Msun \ was
calculated with three orders of magnitude finer surface zoning (to
10$^{-8}$ \Msun) than in HW02.  All models were calculated using the
\kepler\ code and the physics discussed in HW02 and
\citet{Woosley_Weaver_2002}.  One difference with the previous studies
is that mass loss was included in the hydrogenic stars with non-zero
metallicity; however, the mass lost was both small and very uncertain.

Properties of all presupernova stars and their explosions are given in
Table 1.  The major distinction between the zero metallicity and
10$^{-4}$ solar metallicity presupernova stars was that the former
died as compact blue supergiants (BSG) while the latter were red
supergiants (RSG) with radii 10 to 50 times larger. To some extent,
this difference also relies on a particular choice of uncertain
parameters -- in particular primary nitrogen production and mixing --
so not too much weight should be placed on the metallicity of the
model. The production of primary nitrogen was, by design, minimal in
all models. In the zero metallicity series, especially the higher mass
ones, this required reducing semi-convection by a factor of 10 compared
with its usual setting in \kepler\ and turning off overshoot
mixing. Unless this was done, only the 150 \Msun\ model ended up as a
blue supergiant while the other four were red owing to primary
nitrogen production.

For the 150, 175, and 200 \Msun \ stars with 10$^{-4}$ solar
metallicity, both semi-convection and overshoot mixing had their
nominal settings, but red supergiants resulted without primary
nitrogen production. The 225 and 250 \Msun\ stars with 10$^{-4}$
solar metallicity used the reduced mixing - again to avoid large
$^{14}$N production - but also ended up as red supergiants. Since even
a moderate amount of rotation would lead to some mixing between the
helium core and hydrogen envelope in the zero metal stars, and since
the nominal semi-convection and overshoot parameters also lead to
primary nitrogen production, it is likely that a fraction, perhaps
all, of the zero metallicity pair instability stars also die as red
supergiants. This would imply that blue supergiants are a rare
population of supernova progenitors for such massive stars even at
ultra-low metallicity. Our goal here though was to prepare a set of
blue and red supergiant progenitors with a range of masses to explore
how the observable properties of the explosion may depend on the
structure of the hydrogen envelope.

\subsection{Explosion}
\label{sec:explode}

\begin{figure*}
\includegraphics[width=6.0in]{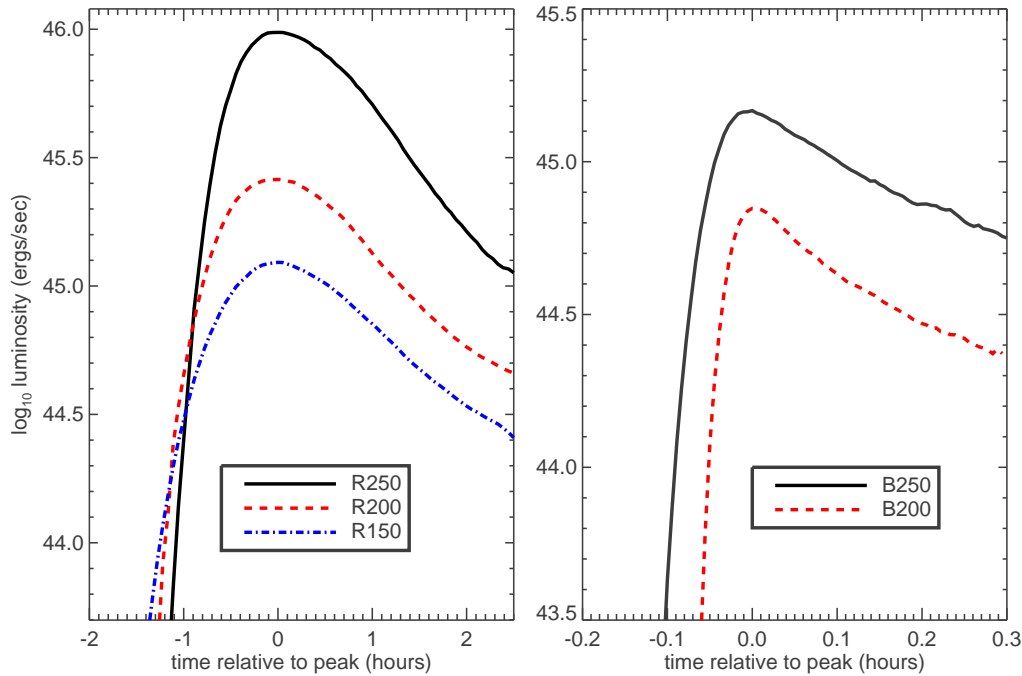}
\caption{Calculated bolometric light curves of the 
shock breakout transient from pair-instability explosions of massive
red-supergiant (left) and blue-supergiant (right) stars. The duration of
the burst, determined by the light crossing time, is significantly
longer for the more extended red supergiant models.
\label{Fig:bo_lc} }
\end{figure*}

The pair instability is triggered in helium cores above about 40
\Msun\ once the temperature in the stellar core exceeds $\sim10^9$ K,
i.e., after helium burning and during carbon ignition.  The creation
of $e^+/e^-$ pairs in the core then softens the equation of state to
below $\gamma < 4/3$ leading to instability and collapse.  For helium
cores below about 135 \Msun\ the collapse is eventually reversed by
explosive nuclear burning, first of oxygen and, for more massive
cores, silicon. Non-rotating helium cores more massive than 133 \Msun\
experience a photo-disintegration instability after silicon burning and
collapse directly to a black hole (HW02).  Between 40 and 60 \Msun,
the pair instability in bare helium cores leads to multiple violent
mass ejections, but does not disrupt the entire star on the first try
\citep{Woosley_PP}. Between 60 and 133 \Msun, helium stars collapse to
increasing central temperature and density, explode with greater
violence, and produce more heavier elements, especially \Nifs\ (see
Table~\ref{tab:models}).

Most of the helium core explosions studied here were taken from the
survey by HW02, but the 100 \Msun\ model was recalculated with finer
surface zoning. A possible point of confusion is whether shocks ever
form in these sorts of helium stars or whether, given their small
radius, the surface remains in sonic communication with the center
throughout the collapse and initial expansion. We find that for the
100 \Msun\ model the outer layers do not participate in the collapse
and, at about 2 seconds after maximum compression, a very strong shock
forms initially about 2 \Msun\ beneath the surface at a radius of $8
\times 10^9$ cm.  Due to the large explosion energy and acceleration
in the steep density gradient, material in the shock reached a speed
of about one-third the speed of light before erupting through the
photosphere.

The explosions of the hydrogenic stars had characteristics, including
kinetic energy and nucleosynthesis, approximately set by the mass of
their helium cores (Table ~\ref{tab:models}). There were, however,
some important differences.  For the helium stars the mass was set at
the beginning of the calculation and held fixed, while in the
hydrogenic stars the helium core grew significantly after central
hydrogen depletion owing to hydrogen shell burning. In some cases it
also shrank due to convective dredge up. At death the helium core thus
had a different composition and entropy than a corresponding helium
star evolved at constant mass.

An even more significant difference of the hydrogenic stars is that
their exploding helium cores encounter the lower density hydrogen
envelope and interact with it hydrodynamically. This has several
important consequences.  First, the expansion of the helium core is
slowed, which produces Rayleigh-Taylor instabilities and mixing. Up
until this point, the explosion had been determined by simple physics
and (assuming no rotation) well represented by a spherical
one-dimensional calculation.  While the mixing can be calculated in a
multi-dimensional code \citep[e.g.,][]{Joggerst_2009a}, it has not
been yet and is parametrized here as in \citet{Kasen_SNII}. Except for
two models that experienced significant fallback (see below), the
mixing has no effect on the nucleosynthesis or energetics and does not
affect the breakout emission. However, the late time spectra may be
somewhat sensitive to mixing.  Figures~\ref{Fig:density} and
\ref{Fig:abun} plot the final density and mixed compositional
structures for a few representative models.

\begin{figure*}
\includegraphics[width=6.0in]{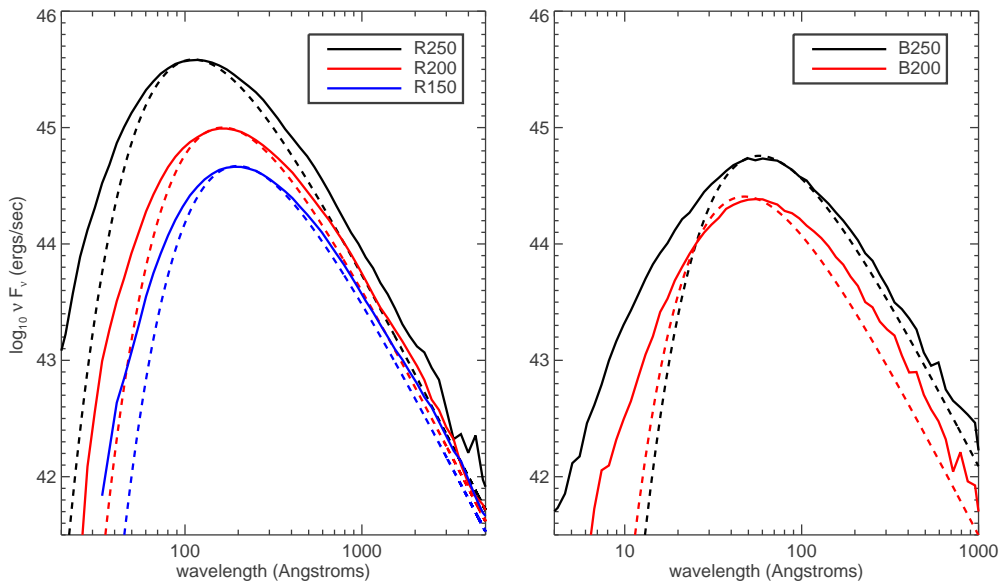}
\caption{Synthetic spectra of the shock breakout transients averaged over the burst peak. 
The time-averaged spectra are the convolution of several blackbodies at different
temperatures, and so are broader than a single blackbody function
peaking at the same wavelength (dashed lines).
Due to the low thermalization opacity, the color temperature is greater than the photospheric temperature
by a factor of $2-3$.
\label{Fig:bo_spec} }
\end{figure*}

In some cases, the deceleration of the helium core by the envelope is
so severe that the star doesn't completely explode. This is
particularly true for the blue supergiant models.  As pointed out by
\citet{Chevalier_1989} and explored by \citet{Zhang_2008} and
\citet{Joggerst_2009b}, braking and fall back occur to a greater
extent in more compact stars. The helium core encounters its own mass
earlier and the reverse shock returns to the center when the density
is still high. In our case, this resulted in models B150 and B175
failing to eject anything other than their hydrogen envelope. Of
course the story does not end there -- these bound helium cores
oscillate awhile, settle down and evolve again. Given their residual
helium masses, models B150 and B175 will probably become pair
instability supernovae again and disrupt entirely on the second
try. They thus represent an extension to higher masses of the
pulsational pair instability supernovae studied by
\citet{Woosley_PP}. Depending upon the timing, the second mass
ejection may produce an {\sl extremely} bright supernova as the very
energetic second supernova plows into the ejecta of the first. This
future evolution is beyond the scope of the present paper, but the
evolution and explosion of blue supergiants with ZAMS masses of 100 -
200 \Msun\ is clearly an area worth further study.

\section{Observable Properties}

\subsection{Shock Breakout}
\label{sec:bo}

In the hydrogenic models, the expansion of the exploded helium core
into the hydrogen envelope drives a radiation dominated shock.  When
this shock approaches the surface of the star, the postshock radiation
can escape in a luminous x-ray/UV burst \citep{
  Klein_Chevalier_1978,Ensman_bo,Matzner_McKee}.  This event occurs at
a distance $\Delta R$ from the stellar surface such that the diffusion
time from the shock front ($t_d \sim \tau \Delta R/c$) is comparable
to the dynamical time for the shock to travel the same distance ($t_e
\sim \Delta R /\vs$).  This implies an optical depth $\tau \approx
c/\vs$ at breakout.  The shock velocity \vs\ scales with $(E/M)^{1/2}$
and for pair SNe is of order $3\times 10^3 - 10^4$~\kms, similar to
that of ordinary core collapse supernovae.  Shock breakout thus occurs
at optical depths of $\tau_b \approx 30-100$.

The surface layers of PI~SNe progenitors are difficult to model; a
proper representation of the atmospheric structure would require a
detailed radiation transport within the stellar evolution code and possibly the inclusion of 3-D
effects.  The limitations of the 1-D stellar evolution models thus
introduce some uncertainty into the breakout predictions.  While the
current models were much more finely zoned than previous calculations,
they still did not fully resolve the optically thin layers of the
star.  They did, however, determine the slope of the steep density
profile at the surface, out to an optical depth of a few.  To extend
the profile into the optically thin region, we fit a power law to the
outermost zones and linearly interpolated and extrapolated the density
structure.  The re-zoned model had 100 zones within the region $\tau <
\tau_b$, and extended to a minimum $\tau$ of $10^{-2}$.

We followed the supernova explosion using the \kepler\ code until the
shock front began to approach the stellar surface ($\tau \approx
10^5$).  The model structure was then mapped into a modified version of the
\texttt{SEDONA} transport code \citep{Kasen_MC} which coupled
multi-wavelength radiation transport to a staggered mesh
1-D spherical Lagrangian hydrodynamics solver. A standard artificial viscosity
prescription was included to damp oscillations behind the shock front.  
We followed the radiation transport
using implicit Monte Carlo methods \citep{Fleck_Cummings} coupled to the
hydrodynamics  in an operator split way.  The transport was treated in a mixed-frame
formalism, in which photon packets were propagated in the inertial frame,
but the opacities and emissivities were computed in the comoving frame, and the proper Lorentz transformations were applied to move between the frames.
The radiation energy and momentum deposition terms in the 
hydro equations were estimated
by tallying the energy and direction of packets moving through each
zone.  
 Further details of the numerical methods in the context of shock breakout 
 will be given in  a separate publication \cite{Kasen_BO}.

\begin{deluxetable*}{ccccccccc}[b]
\tablewidth{6.0in}
\tablecaption{Shock Breakout Transients}
\tablehead
{
\colhead{name}                & 
\colhead{$L_{\rm peak}$ (\ergss)}       & 
\colhead{$\Delta t_h$\tablenotemark{a}}         & 
\colhead{$T_{\rm eff} (10^5)$}        & 
\colhead{$T_{\rm col} (10^5)$\tablenotemark{b}}      & 
\colhead{$\lambda_{\rm p}$\tablenotemark{c}}    & 
\colhead{$E_{\rm p}$\tablenotemark{d}}        & 
\colhead{$E_{\rm tot}$\tablenotemark{e}}    &
\colhead{$E_{> L_\alpha}$\tablenotemark{f}}
}
\startdata                                      
R250  &   9.6$\times 10^{45}$   &   5858  &   1.3   &   3.5   &   82.5   &   0.15   &   5.9$\times 10^{49}$   &   2.2$\times 10^{47}$ \\
R200  &   2.6$\times 10^{45}$   &   6422  &   1.0   &   2.2   &   130.5  &   0.09   &   1.8$\times 10^{49}$   &   1.6$\times 10^{47}$ \\
R150  &   1.2$\times 10^{45}$   &   7051  &   0.9   &   1.7   &   168.5  &   0.07   &   9.3$\times 10^{48}$   &   1.3$\times 10^{47}$ \\
B250  &   1.4$\times 10^{45}$   &   966   &   3.3   &   6.3   &   45.5   &   0.27   &   1.3$\times 10^{48}$   &   6.5$\times 10^{44}$ \\
B200  &   6.7$\times 10^{44}$   &   714   &   3.9   &   7.7   &   37.5   &   0.33   &   5.8$\times 10^{47}$   &   1.9$\times 10^{44}$ \\
\enddata
\tablenotetext{a}{duration of burst, full-width half maximum, in seconds}
\tablenotetext{b}{color temperature}
\tablenotetext{c}{spectral wavelength peak, in \AA}
\tablenotetext{d}{spectral energy peak, in keV}
\tablenotetext{e}{total energy emitted in burst, in ergs}
\tablenotetext{f}{total energy emitted in burst at wavelengths greater than Lyman-$\alpha$, in ergs}
\label{tab:bo}
\end{deluxetable*}

The opacity and emissivity of the models were discretized into 25000 wavelength bins covering the
range $\lambda = 0.1-25000$~\AA, while the output spectra were binned
to a coarser resolution to improve the photon statistics.
The dominant opacity in the supernova shock is electron scattering,
while free-free is the most important absorptive opacity.  Because the
current models had zero or very low metallicity, we ignored bound-free
and line opacity.  At wavelengths near the blackbody peak, the ratio
of free-free opacity to total opacity is typically small $\epsilon_a
\la 10^{-4}$.  In the postshock region, Compton up-scattering may then
become an important means of energy exchange between matter and
radiation \citep{Weaver_shock}.  On average, the fractional change of
a photon's energy in a single Compton scattering is $\epsilon_c = 4 k
T/m_e c^2$, which gives an effective $\epsilon_c = 6 \times 10^{-5}$
at $T = 10^5$~K.  Comptonization will significantly alter the
radiation spectrum after scattering through an optical depth $\tau_c
\ga \epsilon_c^{-1/2}$ which, for higher temperatures, may be less
than the thermalization depth to free-free opacity $\tau_a =
\epsilon_a^{-1/2}$.  In our calculations, we therefore approximated
the Compton effects by taking a fraction $\epsilon_c$ of the Thomson
opacity to be thermalizing.  This grey absorptive component was added
to the wavelength dependent free-free opacity. Clearly a direct
treatment of inverse Compton scattering  is needed to more accurately
predict the spectrum of the breakout burst; nevertheless, our
treatment of the full non-grey radiation transport offers some
advance over previous numerical calculations.

Once a shock front reaches the steep outer layers of the star, it
accelerates down the steep density gradient.  For model R250, the
shock velocity reaches $v_s \approx 1.5 \times 10^4$~\kms\ around
breakout.  The temperature of the post-shocked gas is determined by
the jump conditions $a T_s^4/3 = \rho_0 \vs^2/(\gamma +1)$, where
$\rho_0$ is the density of the pre-shocked material and $\gamma = 4/3$
is the adiabatic index for a radiation dominated gas.  Near the
surface, the RSG models had densities of $\rho_0 \sim 10^{-11}$~\gcc\
which gave typical postshock temperatures of $T_s \approx 5 \times
10^5$~K.  The BSG models, being more compact, had higher densities at
the surface and hotter temperatures, $T_s \approx 10^6$~K.


Figure~\ref{Fig:bo_lc} plots the calculated bolometric light curves of
the breakout transients.  The key observable properties are listed in
Table~\ref{tab:bo}.  These calculations included the proper light
travel time effects, which primarily determined the duration of the
observed burst, $\Delta t \approx R_0/c$.  The RSG models lasted 1 to
2 hours, and reached peak luminosities of $L_{\rm peak} =
10^{45}-10^{46}$~\ergss .  The BSG models, being more compact, had
briefer transients lasting $0.1$~hours, with somewhat lower peak
luminosities, $L_{\rm peak} \approx 10^{44}$~ergs/sec.  These values
can be compared to the breakout in an ordinary Type~IIP supernova
explosion: $L_{\rm peak} = 5 \times 10^{44}$~\ergss, $\Delta T =
0.5$~hours.   As it turns out, the shock velocities
and post-shock energy densities of PI~SNe are comparable to that of
Type~IIP supernovae, but the breakout emission can be significantly
brighter due to the larger stellar radii.  

Figure~\ref{Fig:bo_spec} shows synthetic spectra of the bursts,
averaged over the peak of the breakout light curve.  The RSG spectra
peak at wavelengths $\lambda_p = 80-170$~\AA\ ($\sim 0.07-0.15$~keV),
while the BSG spectra peak near $37-45~\AA$ ($\sim 0.3$~keV).  These
peak wavelengths are comparable to that of ordinary Type~IIP breakout
($\lambda_p \approx 100$~\AA).  The spectra are reasonably
approximated by a blackbody, but with excess emission at both high and
low frequencies.  This is because the time-averaged spectra represent
a convolution of several blackbodies at different temperatures.  

Because the opacity in the atmosphere is strongly scattering
dominated, radiation is thermalized at optical depths $\tau \approx
\epsilon_a^{-1/2}$ well below the photosphere.  The emergent spectrum
is thus characterized by the higher temperature of these deeper
layers.  Defining the color temperature $T_{\rm col}$ of the mean
spectrum as the temperature of a blackbody peaking at the same
wavelength, we find that the observed $T_{\rm col}$ is a typically a
factor 2-3 times higher than the effective temperature at the
photosphere (Table~2).  A similar effect
was noted by \cite{Ensman_bo} for models of SN~1987A.
In addition, recent analytic work by \cite{Katz_2010}
has shown that when the shock velocity is relatively high, $\ga 0.1c$ (and when 
the thermalization by Compton scattering is treated properly)
non-equilibrium effects may lead to a significantly harder
non-thermal component in the spectrum. 
In the BSG models, the  shock velocities do in fact reach  $ \sim  0.1 c$ 
and such a non-thermal component could modify the predicted spectra
shown here. The
the RSG models, on the other hand, have lower shock velocities and are likely
not strongly affected.

Following shock breakout, radiation continues to diffuse out of
the expanding, cooling ejecta.  This leads to a longer lasting 
emission that may be visible at optical wavelengths for several weeks.  Such an
 initial thermal component to the light curves 
is discussed in the following section.
The extreme luminosity of the breakout bursts themselves, and their short
durations, make them appealing transient to search for in
high-redshift surveys.  We consider their detectability in
Section~\ref{sec:detect}. 

\subsection{Light Curves} \label{sec:lc}

\begin{figure*}
\includegraphics[width=7.0in]{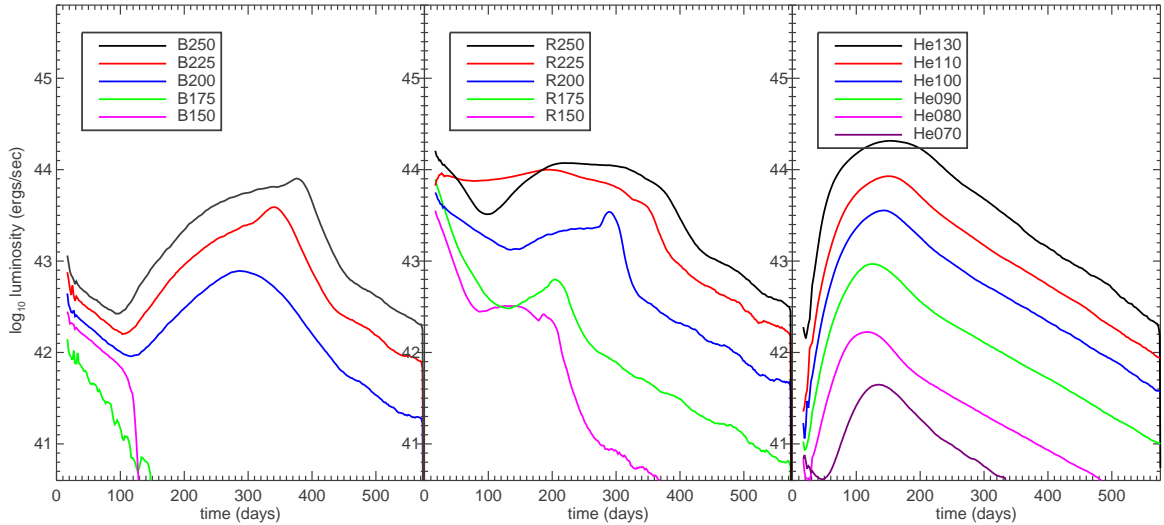}
\caption{Bolometric light curves of the full set of models representing the 
explosion of blue supergiant stars (left panel), red supergiant stars
(middle panel) and bare helium cores (right panel).  The more massive
stars have brighter and longer lasting light curves.
\label{Fig:bol} }
\end{figure*}

\begin{figure}
\includegraphics[width=3.5in]{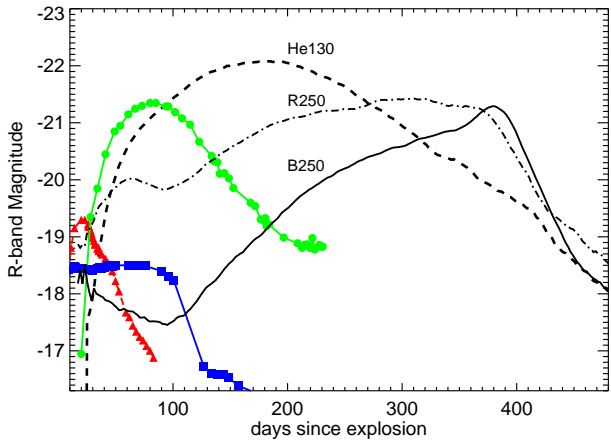}
\caption{Synthetic R-band light curves (at $z=0$) of 
bright PI~SN models -- R250 (dashed-dot), B250 (solid), and He130
(dashed) -- compared to observations of a normal Type~Ia supernova
SN~2001el \citep[red triangles,][]{Kris_01el} a normal Type~IIP
supernova SN~1999em \citep[blue squares,][]{Leonard_99em_lc} and the
over-luminous core-collapse event SN~2006gy \citep[green circles,][]{Smith_06gy}.
 \label{Fig:RLC}}
\end{figure}

\begin{figure*}
\includegraphics[width=7.0in]{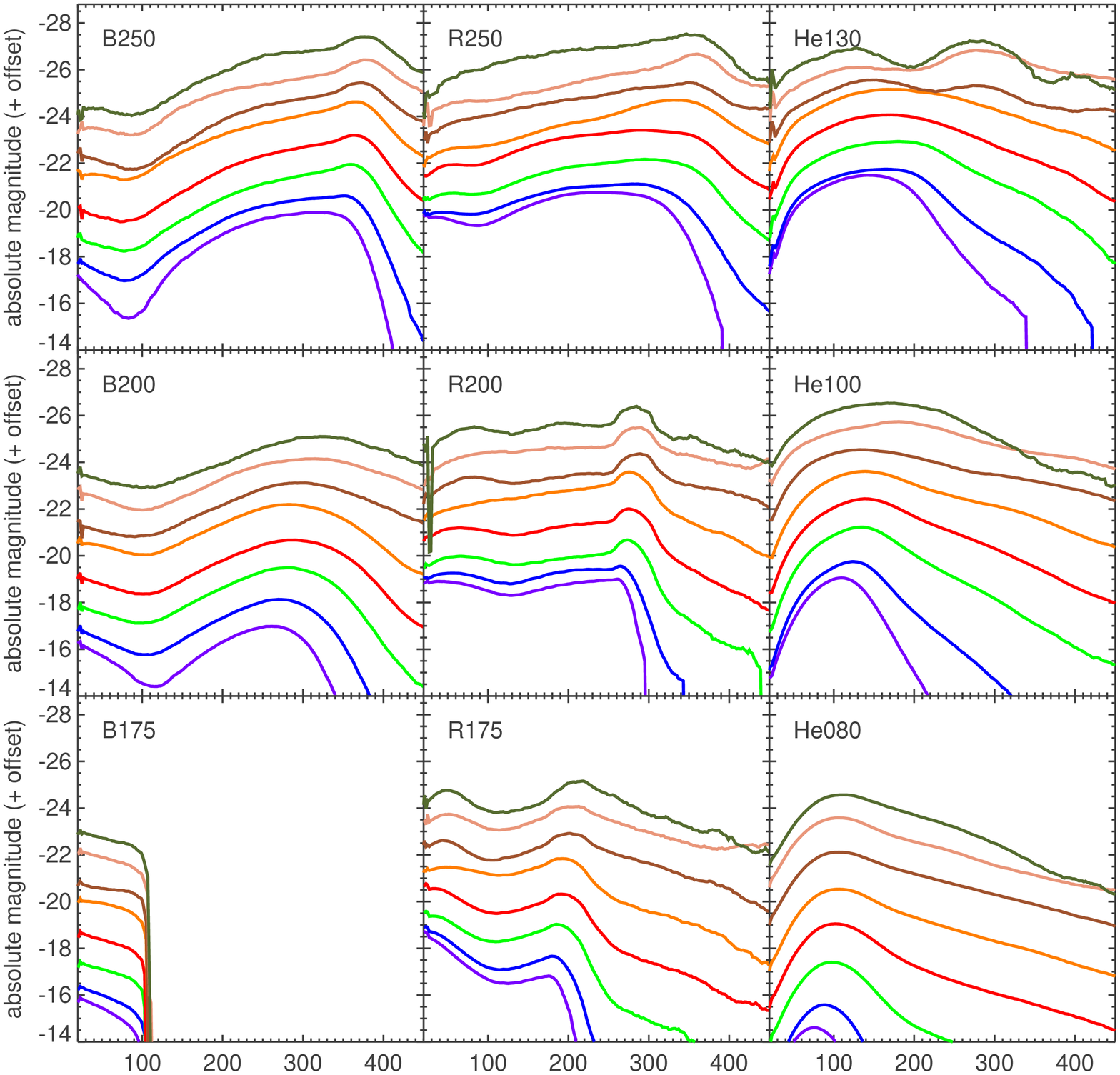}
\caption{Broadband light curves of PI~SN models.  The
light curves displayed in each panel are, from bottom to top,
UBVRIJHJK.  The y-axis corresponds to the B-band light curve, while
the other light curves are offset by 1 magnitude from the one above or
below.
\label{Fig:broadband} }
\end{figure*}

Following shock breakout, the luminosity of PI~SNe can be powered by
three different sources: (i) The diffusion of thermal energy deposited
by the shock; (ii) the energy from the radioactive decay of
synthesized \Nifs; or (iii) the interaction of the ejecta with a dense
surrounding medium.  The internal energy suffers adiabatic losses on
the expansion timescale $t_{ex} = R_0/v$, so source (i) will be most
significant for stars with large initial radius $R_0$.  Interaction
has not been included in the models discussed here, but as already
mentioned, can be very significant for stars which undergo pulsations
before exploding completely.

We calculated light curves of the explosion models using the
\texttt{SEDONA} code \citep{Kasen_MC}.  The initial density,
composition, and temperature structures were taken from the \kepler\
calculations extended into the nearly homologous expansion phase
($\sim 10$~days after explosion). The energy deposition from \Nifs\
decay was followed using a multi-wavelength transport scheme treating
the emission, propagation, and absorption of gamma-rays.  For the
optical radiation transport, the opacities used included
electron-scattering, bound-free, free-free, and the aggregate effect
of millions of Doppler broadened line transitions treated in the
expansion opacity formalism \citep{Eastman_93}.  Atomic level
populations were calculated assuming local thermodynamic equilibrium,
typically a reasonable approximation for supernovae in the
photospheric phase.

The resulting model light curves (Figure~\ref{Fig:bol}) span a wide
range of luminosities and durations.  The more massive explosions are
bright for over 300 days, with luminosities exceeding
$10^{44}$~ergs/sec. The long duration reflects the timescale for
photons to diffuse through the optically thick supernova ejecta.  In a
homologously expanding medium, the effective diffusion time scales as $t_d \sim
\kappa^{1/2} M_{\rm ej}^{3/4} E^{-1/4}$, where $\kappa$ is the
effective mean opacity \citep{Arnett_80}.  Model He130, for example,
is nearly 100 times as massive and energetic as a SN~Ia, and thus has
a light curve $\sim 10$ times as broad.  The slow release of energy
moderates the emergent luminosity, so that despite making over 60
times as much \Nifs\ as a typical SNe~Ia, model He130 is only $\sim
10$ times brighter than one at peak.

The morphology of the model light curves depends on the envelope of
the progenitor star.  The RSG models resemble Type~II plateau
supernovae (SNe~IIP) light curves, though with luminosities and
durations $\sim 3$ times greater.  The initial luminosity is powered
by the thermal energy in the extended hydrogen envelope.  Over time,
as the ejecta cool, a hydrogen recombination front propagates inward
in mass coordinates, which increases the ejecta transparency due to
the elimination of electron scattering opacity.  This effect regulates
the release of the thermal energy \citep{Grassberg_1976, popov,
  Kasen_SNII}.  In the inner regions, heating from radioactive decay
delays recombination and causes the light curve to rise to a peak at
around $200-300$~days.  Eventually, the transparency wave reaches the
base of the hydrogen envelope, after which recombination proceeds much
more rapidly through the heavier element core.  After a rapid release
of the remaining internal energy, the light curve drops off sharply
and follows directly the radioactive energy deposition rate.

\begin{figure*}
\includegraphics[width=7.0in]{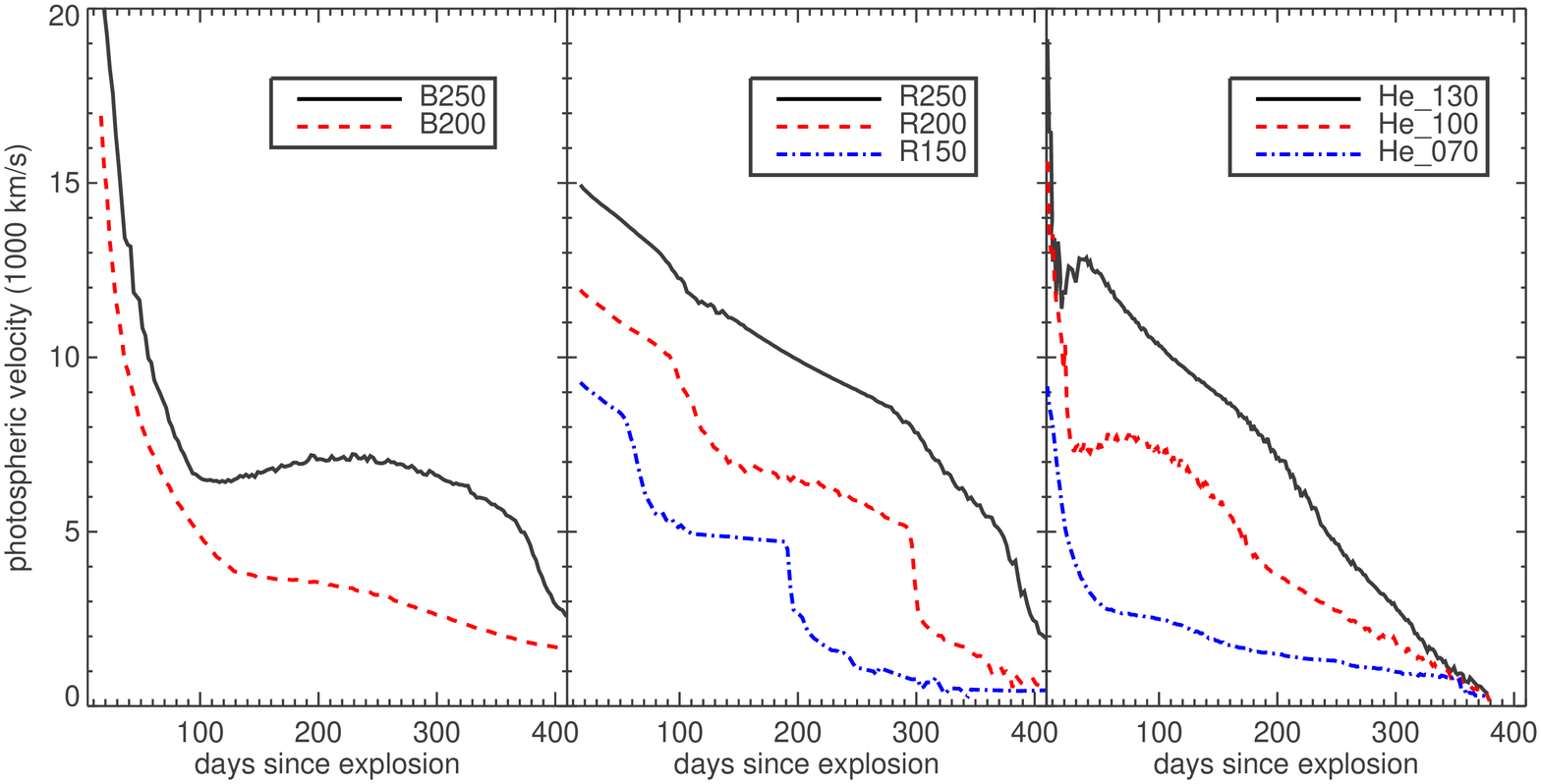}
\caption{Time evolution of the  velocity measured
at the electron scattering photosphere for several PI~SN models.  In
the BSG and helium core models, the photosphere initially recedes
quickly as the outer layers recombine and become transparent.  The
rise in photospheric velocity seen in some models reflects the outward
diffusion of radioactive energy which may heat and reionize the
external layers.
\label{Fig:vphot} }
\end{figure*}

\begin{figure}
\includegraphics[width=3.5in]{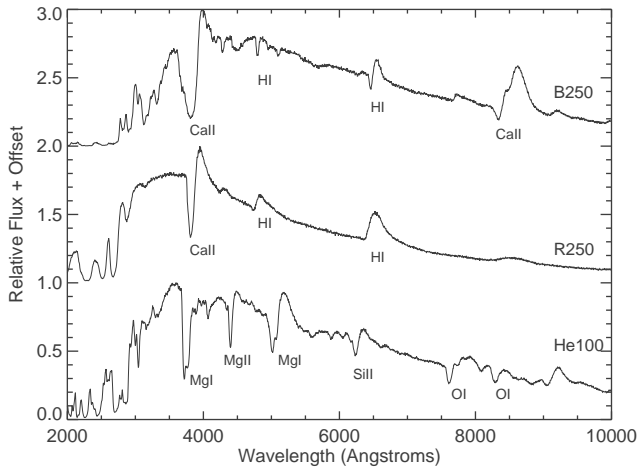}
\caption{Spectra of select models at the peak of the light curve.  
The important line absorption features are marked.
\label{Fig:max_spec} }
\end{figure}

The light curves of the BSG models have a dimmer initial thermal
component, due to the relatively smaller radii of the progenitors.  In
this respect they resemble scaled up versions of SN~1987A.  In models
B200, B225, and B250, the light curves rise to a bright \Nifs\ powered
peak at $\sim 300$ days after explosion.  The less massive events
(B150 and B175), which failed to eject any \Nifs, only show the
initial thermal light curve component lasting $\sim 150$~days, and
more closely resemble typical SNe~IIP.

The helium core PI~SNe models, being the most compact progenitors,
lack a conspicuous thermal light curve component altogether.  The
lower total mass leads to relatively briefer light curves, peaking
around 150~days after explosion.  Model He130 ($\sim 40$~\Msun\ of
\Nifs) reaches an exceptional peak brightness of $2 \times
10^{44}$~\ergss\ at around 180 days.  Model He70, on the other hand,
proves that despite being massive and energetic, not all PI~SNe are
bright.  This explosion produced only $0.02$~\Msun\ of \Nifs\ and the
light curve reached only $3 \times 10^{41}$~ergs/sec, rather
sub-luminous for a supernova.

Figure~\ref{Fig:RLC} compares the synthetic R-band light curves of
representative PI~SN models with observations of a typical Type~Ia
supernova \citep[SN~2001el,][]{Kris_01el}, a typical Type~IIP
supernova \citep[SN~1999em,][]{Leonard_99em_lc}, and SN~2006gy, one of
the most the luminous core-collapse SNe discovered
\citep{Smith_06gy}.  The later has been suggested to be a pair
instability supernovae, however the predicted model light curve
durations are seen to be too long even for this event by a factor of
several.  Another possibility is that SN~2006gy was an example of a
pulsational-pair instability supernova \citep{Woosley_PP}.

Figure~\ref{Fig:broadband} shows the multi-color optical and
near-infrared light curves for the models.  At bluer wavelengths, the
light curves generally peak earlier and decline more rapidly after
peak.  This reflects the progressive shift of the spectral energy
distribution to the red over time.  This shift is due not only the
decrease in effective temperature, but also to the increase in line
opacity from iron group elements of lower ionization stages (Fe~II and
Co~II), which blankets the bluer wavelengths.

The brightest helium core models such as He130 display a pronounced
secondary maximum in the infrared light curves.  This feature is
similar to what is observed in Type~Ia supernovae and the physical
explanation is essentially the same \citep{Kasen_IRLC}.  When the
temperature in the ejecta drops below $\sim 7000$~K, doubly ionized
iron group elements begin to recombine.  As the infrared line
emissivity is much greater for singly ionized species (Fe~II and
Co~II) the flux can be more efficiently redistributed from bluer to
redder wavelengths, leading to an increase in the infrared luminosity.
The secondary maximum is therefore more prominent in models with large
abundances of iron group elements.  Detection of a secondary maximum
in an observed supernova would provide strong evidence that the
explosion did indeed synthesize large amounts of \Nifs.  The lack
of a secondary maximum does not necessarily rule out the presence
of substantial \Nifs, as strong radial mixing of the nickel can sometimes smear
out the two bumps \citep{Kasen_IRLC}.

\subsection{Spectra}
\label{sec:spectra}

\begin{figure}
\includegraphics[width=3.5in]{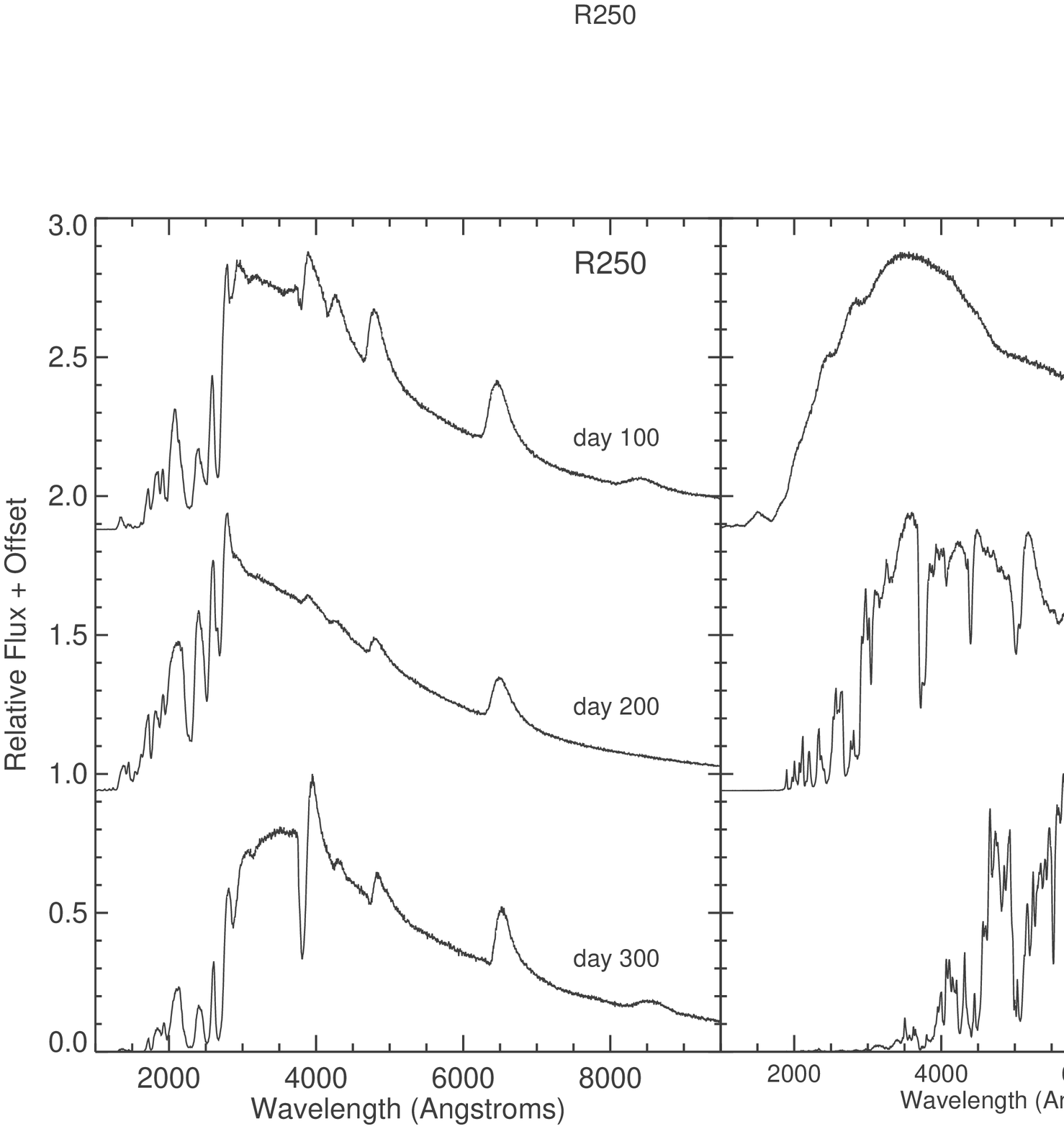}
\caption{Spectral evolution of model R250 (left panel) and 
model He100 (right panel).  Over time, additional line features
appears as the photosphere recedes into layers of burned material.
\label{Fig:spec_series} }
\end{figure}

\begin{figure}
\includegraphics[width=3.3in]{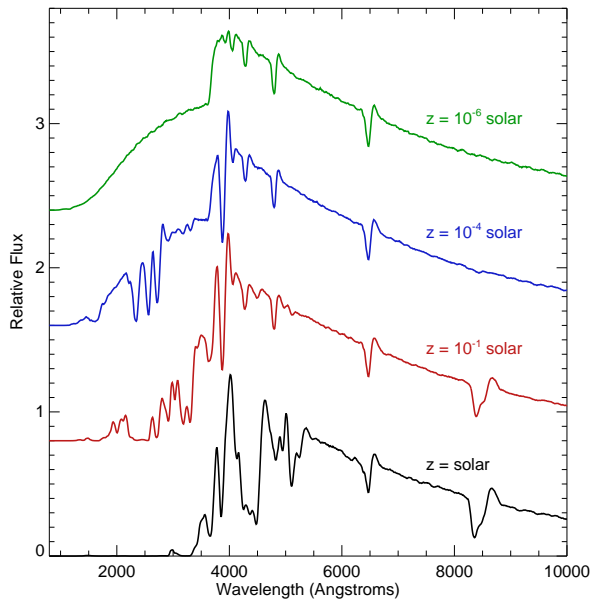}
\caption{The day 100 spectrum of model R150 shown for different values of
metallicity in the hydrogen envelope, from top to bottom $Z = 10^{-6},
10^{-4},10^{-2},$ and $1$ times solar.  The metallicity can be probed
observationally by examining  metal line absorption features or
the flux in the ultra-violet.
\label{Fig:met} }
\end{figure}

\begin{figure}
\includegraphics[width=3.3in]{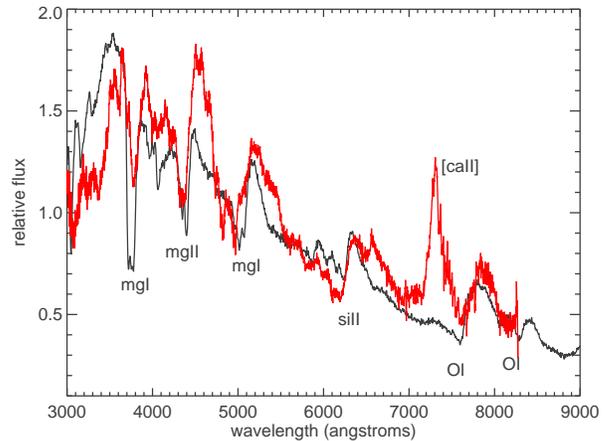}
\caption{The spectrum, observed near peak, of SN~2007bi \citep[red line][]{gal_yam_2009} 
compared to the
synthetic spectrum of our model He100 at 100 days after explosion (black line).
Line identifications for the model are marked.  Most of the major observed line
features are reproduced by the model, except for the forbidden  Ca~II line emission at 7300~\AA\ which 
is the result of non-equilibrium effects not included in the calculations.
\label{Fig:sn07bi}}
\end{figure}

The spectra of the PI~SN models (Figures~\ref{Fig:max_spec} and
\ref{Fig:spec_series}) resemble those of ordinary SNe, with P-Cygni
line profiles superimposed on a pseudo-blackbody continuum. Because of
the low abundance of metals in unburned ejecta, many of the familiar
line features are weak or missing in the early time spectra.  For
example, at maximum light models R250 and B250 show only features from the hydrogen
Balmer lines and calcium in their spectrum.  However, at later times, when the
photosphere has receded into layers of burned material, other line
features appear.

The maximum light spectrum of the bare helium core model He100 is dominated by lines from freshly
synthesized intermediate mass elements, and resembles a Type~Ic SN
with lines due to Mg~I, Mg~II, Si~II, Ca~II and O~I.  After peak, the
spectrum shows more features from the iron group elements in the
\Nifs-rich core.  Helium lines are not present at any epoch, as the
envelope temperatures are too low to thermally excite the lower atomic
levels of the optical transitions.  However, if \Nifs\ is mixed out
into the helium rich layers, non-thermal excitation by radioactive
decay products could generate significant helium line opacity
\citep{Lucy_He}.

Although PI~SNe are highly energetic events, their large ejected
masses imply only moderate characteristic velocities: 
$v \sim (2 E_{\rm K}/M_{\rm
  ej})^{1/2} \sim 5000$~\kms, about half that typical of SNe~Ia
and several times less then the broad lined Type~Ic SNe that have been
associated with gamma-ray bursts \citep{Galama_98bw}.
Figure~\ref{Fig:vphot} show the time evolution of the velocity
measured at the electron scattering photosphere. In the BSG and helium
core models, the photosphere initially resides in the outer, high
velocity layers of ejecta, but recedes quickly as these layers
recombine and become transparent. Eventually, the photosphere settles
in the inner regions of burned material, where radioactive energy
deposition maintains the ionization state. Interestingly, the
photospheric velocity in model B250 in fact {\it increases} for some
time period as the radioactive energy diffuses outward and reionizes
the hydrogen envelope.  In contrast, the decline in photospheric
velocity in the RSG models is gradual until the recombination front
reaches the base of the hydrogen envelope, after which it drops off
sharply.

Certain features in the spectra of PI~SNe could, if observed, offer
direct confirmation that the progenitor star was of low metallicity.
The metallicity has two principle spectroscopic effects
(Figure~\ref{Fig:met}).  First, for metallicities $Z \geq 10^{-4}$,
some absorption lines from intermediate mass elements are noticeable,
for example those of Ca~II H\&K (near 3800~\AA) and the IR triplet
(near 8300~\AA).  Second, higher metallicity leads to a significant
reduction in the ultraviolet (UV) flux, due to the iron group line
blanketing at bluer wavelengths.  Similar metallicity effects have
been noted in the spectral modeling of Type~IIP SNe
\citep{Baron_93W,Dessart_TypeII} Spectroscopic or rest frame UV
observations of PI~SNe may, together with modeling, constrain the
metallicity of the progenitor star.  This assumes, of course, that the
supernova has been observed early enough that we are seeing layers of
ejecta unaffected by the explosive nucleosynthesis.

To draw some comparison to observations, Figure~\ref{Fig:sn07bi} shows the near maximum light spectrum of one model, He100, compared to the most promising candidate to date for an observed pair supernova, SN~2007bi.  \cite{gal_yam_2009} has previously shown that the light curve
of this supernova is a good match to model He100.  On the whole, the correspondence of the spectra is also rather good, especially considering that the model has not been tuned to match the data.  Most of the major line features are reproduced, in particular the unusually prominent Mg~I and Mg~II lines.  The ejecta velocities (as evidenced by the blueshift of the absorption features) are also roughly in agreement.  
The model fails to reproduce the forbidden  Ca~II line emission at 7300~\AA, but this feature results from non-equilibrium effects which are not included in our calculations.

\section{Detectability}
\label{sec:detect}

\begin{figure}
\includegraphics[width=3.5in]{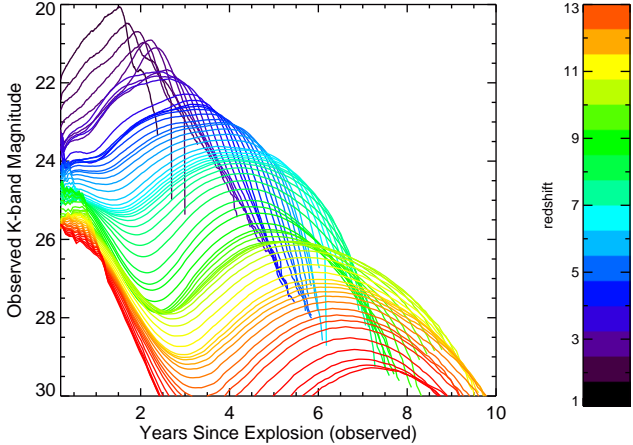}
\caption{Observer frame K-band light curve of model R250 as a function of redshift. 
The effects of cosmological redshift, dimming, and time-dilation have
all been included.  For $z>7$, one observes in the rest-frame UV, and
the initial thermal component of the light curve is brighter than the
later radioactively powered peak.
\label{Fig:det_R250} }
\end{figure}

\begin{figure}
\includegraphics[width=3.5in]{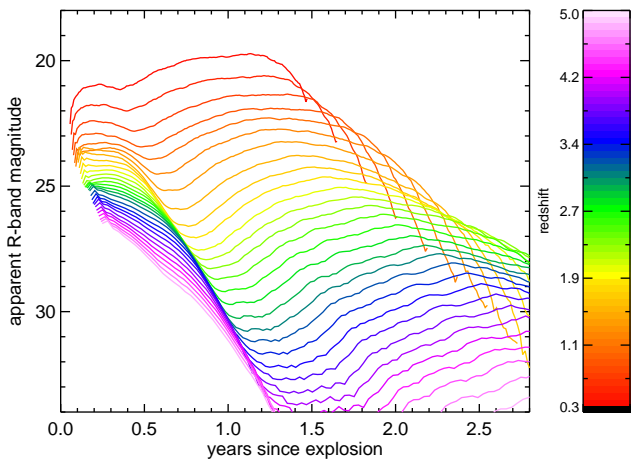}
\caption{Observer frame R-band light curve of model R250 as a function of redshift. 
The effects of cosmological redshift, dimming, and time-dilation have
all been included. 
\label{Fig:det_R250_R} }
\end{figure}

Some PI~SNe should be observable out to large distances, but the
effects of cosmological redshift and time-dilation effects will
significantly affect the shape and luminosity of the observed light
curve.  Figures~\ref{Fig:det_R250} and \ref{Fig:det_R250_R} demonstrate
these effects for model R250 in the K- and R-bands, respectively.  The radioactively powered optical
emission, which lasts for hundreds of days, emits very little flux
at rest wavelengths $\lambda < 2000$~\AA.  Thus, to follow
objects at $z \ge 5$ it is best to observe at infrared wavelengths.
Alternatively, one could search for the brief and very blue emission
from shock breakout.

\begin{figure}
\includegraphics[width=3.5in]{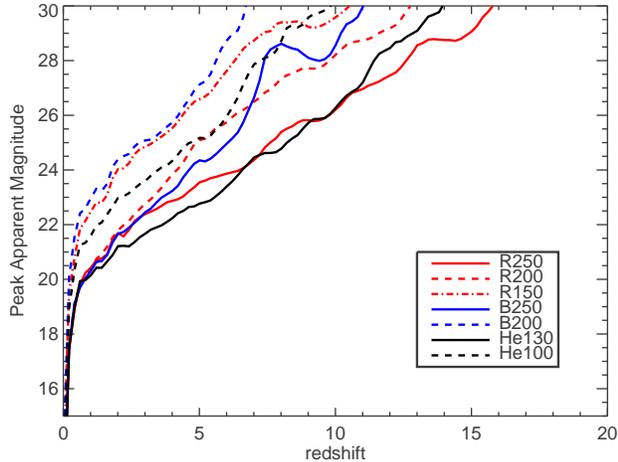}
\caption{Detectability of  pair instability supernovae
light curves.  The figure plots the observer-frame peak K-band magnitude of
several models as a function of redshift.
\label{Fig:Kdetect} }
\end{figure}

\begin{figure}
\includegraphics[width=3.5in]{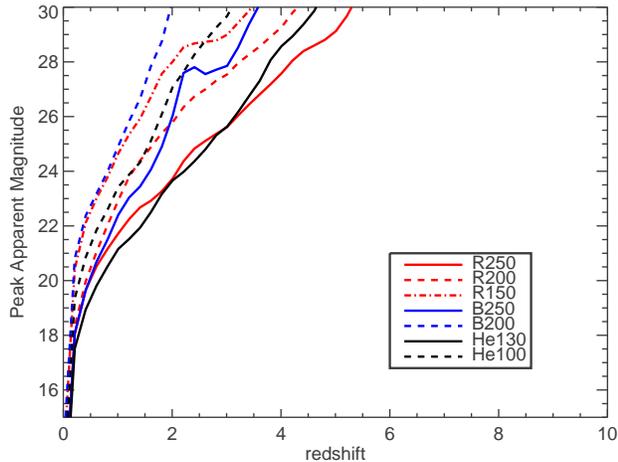}
\caption{Same as Figure~\ref{Fig:Kdetect}, but for the R-band.
\label{Fig:Rdetect} }
\end{figure}

In Figures~\ref{Fig:Kdetect} and \ref{Fig:Rdetect} we plot the
observer frame R- and K-band peak magnitudes of several PI~SN light
curves (starting $t > 10$~days after explosion in the rest-fame) as a
function of redshift.  Future R-band surveys, such as with the Large
Synoptic Survey Telescope, will routinely reach limiting magnitudes of
$M_R \sim 24.5$, which would detect or place constraints on the
brightest PI~SNe out to redshift of $z \sim 2$.  Deeper searches ($M_R
\sim 28$) would reach redshifts $z \sim 5$, while a similar K-band
search would probe redshifts of $z \sim 10$ and beyond.  Space based
observations at wavelengths longer than K-band could potentially see
to even greater distances.  The long duration of the PI~SN light
curves -- greatly prolonged by the $(1+z)$ cosmological time dilution
factor -- poses a challenge for detecting them as transients.  At $z
\sim 7$ the light curve of a PI~SN can last 1 to 5 years in the
observers frame (Figures~\ref{Fig:det_R250} and \ref{Fig:det_R250_R}). 
Deep exposures over long
time baselines would be needed to discover and follow such an event.

An alternative way of discovering PI~SNe would be to search for the
short-lived shock breakout transient.  At $z = 1$, the observed
burst would last a few hours, and at $z=10$ about a day. 
 The spectrum, which peaks in the
rest-frame at $\lambda \approx 80-170$~\AA, would be seen on the
Rayleigh-Jeans tail.  Unfortunately, neutral hydrogen along the line
of sight likely absorbs all radiation shortwards of the Lyman alpha
line at 1215~\AA, which is the bulk of the radiation emitted in shock
breakout.  However, the radiation at longer wavelengths can be
significant and may still be visible.

\begin{figure*}
\includegraphics[width=7.0in]{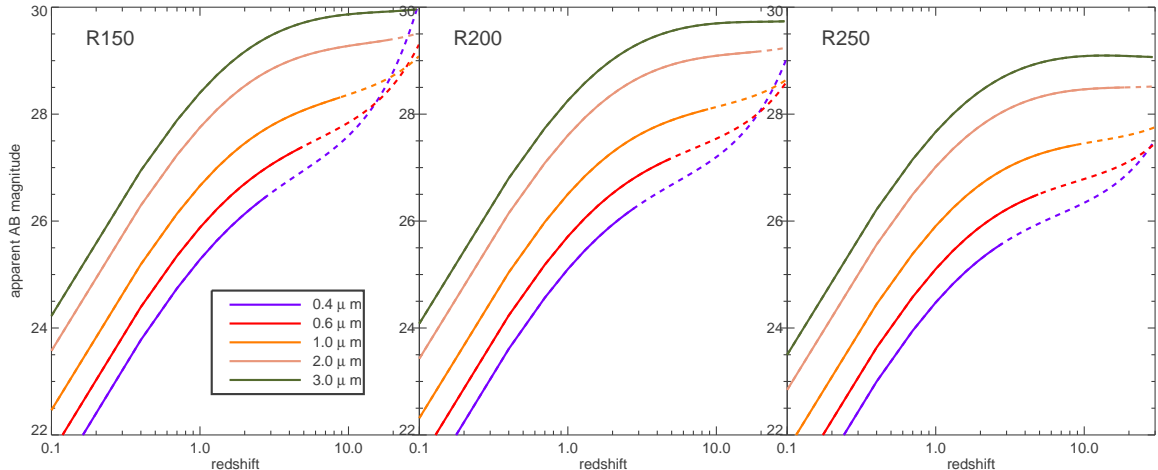}
\caption{Detectability of the shock breakout transient from pair
  instability supernovae as a function of redshift.  The figure plots
  the observed AB magnitude (for different wavelength bands) of three
  red-supergiant models at the peak of the bolometric breakout light
  curve.  (Note an AB magnitude of 28.9 corresponds to $10~$nJ.)
  Solid lines denote the redshift range for which the rest frame flux
  is redder than $1216~\AA$ (the Lyman alpha line); at higher
  redshifts, the flux is likely absorbed by intervening
  neutral hydrogen (dashed lines).  The breakout transients of the
  blue supergiant models will be significantly dimmer than the models
  shown here, as the luminosity is lower and the 
spectrum peaks at much lower wavelengths.
  \label{Fig:bo_detect} }
\end{figure*}

Figure~\ref{Fig:bo_detect} plots the observed AB magnitudes (in
different wavelength bands) of the RSG models at the peak of the
bolometric breakout light curve.  At redshifts $z > 3$ the curves
flatten out as flux is redshifted into the observed band and
compensates for the greater distance.  An all sky optical survey like LSST
which reached magnitude $R \approx 24.5$ per image
could potentially catch shock breakout to redshifts of $z \approx 1$.
In principle, deeper optical imaging in select fields could detect breakout to much
higher redshift, but  in practice, the Lyman limit restricts the optical detectability to $z
\sim 6$.  

Near-infrared surveys could avoid the problem of hydrogen absorption.
A facility like JWST with $\la 10$~nJ ($M_{ab} \sim 29$) sensitivity in the
wavelength region $2-5\mu$ has the capability of detecting 
shock breakout out at redshifts of $z \approx 10-20$. However, the
small field of view of JWST, combined with the expected low rates
of PI~SNe, suggest that it will be unlikely  to  discover a PI~SNe
in this manner. A wide-field imager with infrared capabilities (e.g., WFIRST or EUCLID) 
may fare better for discovering shock breakout at the highest redshifts.

The rates of PI~SNe, in either the nearby or distance universe, are
very uncertain.  \cite{scano_PPSN} (see also \cite{SSF3}) provided one
estimate of the rate as of one PI~SN per 1000 solar masses of metal free stars,
with the metal free stars themselves forming, at all redshifts, at only 1\% of the total star formation rate.  
Observationally, the detection of at least on promising candidate pair~SN (SN~2007bi) suggests a rate
in the local universe of $\la 10^{-4}$ of the core collapse rate \citep{Quimby_09,gal_yam_2009}. 
More detailed investigation of the PI~SN
rate will be considered elsewhere (Pan et al., in prep) but in any case it should
be kept in mind that PI~SNe are rare compared to other supernovae, and will therefore be difficult to discover in bulk.

\section{Summary and Conclusions}

We have surveyed the spectra, multi-color light curves, and shock
breakout transients of pair instability supernova models throughout the mass
range in which the star suffers complete disruption.  Three
varieties of progenitor stars were explored: i) helium cores that may
have resulted when very massive stars either lost their envelopes to
some sort of non-radiative mass loss (as in Eta Carina) or to a binary
companion; ii) blue supergiants still retaining most of their hydrogen
envelope; and iii) red supergiants with enormous radii. The frequency
and distribution with mass and metallicity of ii) and iii) depends on
uncertain parameters of semiconvection and overshoot; in general it
seems that the red supergiants are a more common outcome of the
stellar evolution than blue supergiants.

Though PI~SNe are often considered to be categorically bright events,
in fact it is only the most massive stars ($M > 200$~\Msun) that are
extraordinary in terms of luminosity.  The lower mass objects have
peak brightnesses comparable to normal core collapse events $L =
10^{42}-10^{43}$~ergs, though the light curves last for much longer.
For a sensible initial mass function, it is likely that these dimmer
events are actually the more representative PI~SNe.
An extended light curve, rather than an extreme luminosity,
 may therefore be the most relevant
signature of the pair instability events.

The models illustrate how the different classes of progenitor stars
can be distinguished observationally by the light curve
morphology. The red supergiant explosions have light curves with long
duration plateaus, similar to ordinary Type~IIP supernovae but with
luminosities and durations $\sim 3$ times greater.  The light curves
of the blue-supergiant explosions more closely resemble SN~1987A, with
a brief initial thermal component and a late, prominent \Nifs\ powered
peak.  The bare helium core model light curves resemble very
long lasting Type~Ib/Ic supernovae.

The exploding cores of the more compact blue supergiant progenitors
are braked extensively by hydrodynamic interaction with their
envelopes.  For the lighter stars considered (models B150 and B175)
the system failed to completely explode on the first try, and only the
outer envelope of the star was ejected.  The class of pulsational pair
instability may thus have a larger range than stated by
\citet{Woosley_PP}.  A secondary explosion of the helium core, and its
subsequent collision with the ejected hydrogen envelope, could then
lead to an extremely luminous supernova event.

Despite the large explosion energy of PI~SNe, the expansion
velocities, as measured from the Doppler shift of spectral lines, are
rather low $v \approx 5000 - 10000~\kms$.  The spectra of helium core
explosions resemble Type~Ic supernovae, and in particular are
distinguished by prominent lines of Mg~I and Mg~II.  The spectra of
hydrogenic models appear in most respects like ordinary Type~II
supernovae, however the absence of certain metal line features (e.g.,
the Ca~II IR triplet) and the bright ultraviolet emission (due to the
reduced iron group line blanketing) may provide signatures that the
stellar envelope was indeed of very low metallicity.

The recently  publicized SN~2007bi has been suggested to be the first
convincing detection of a PI~SN.  \cite{gal_yam_2009} compared the observed
light curve of SN~2007bi to the models
presented here, and found very good agreement with a helium core explosion
of mass 100-110~\Msun. We have shown here that the spectrum of the helium core model
is also in reasonable agreement with the observations, and in particular with the presence
of strong magnesium lines.  SN~2007bi is therefore
a compelling candidate for a PI~SN, however other
possible scenarios have been suggested.  \cite{Moriya_2010}  have shown that the light curve could be explained as well by the core collapse of a massive 43~\Msun\ C/O core.  Another possibility  is that the light curve was powered not by \Nifs\ decay, but by energy injection from a highly magnetized remnant neutron star \citep[a magnetar][]{kasen_bildsten,Woosley_09}).  The distinguishing feature of the pair~SNe model is its exceptionally long rise time, but unfortunately the observations of SN~2007bi lacked data before the peak .  At present, one appealing feature of the pair~SNe model for SN~2007bi is that it correctly predicts the explosion energy, given the progenitor star mass.  In the other models, the energy is a free parameter that must be input by hand to match the observed light curves and spectra.  

Other luminous observed supernovae, such as SN~2006gy \citep{Smith_06gy}, SN~2005ap \citep{Quimby_05ap}, and SN~2008es \citep{Gezari_2009} have occasionally been suggested
to be examples of PI~SNe.  However, the light curve duration of these events ($\sim 50-100$ days) are simply too short to be explained by the classic \Nifs\ powered pair explosions explored here, which all last of order $\sim 300$~days.  The pulsational pair scenario, which occurs for stellar masses just below those considered here, provide a possible explanation
for these observations, as does magnetar energy injection.

It is often lamented that pair~SNe cannot be discovered as transients at high redshift -- their intrinsically slow ($\sim 1$ year) light curve evolution, prolonged by a $(1+z)$ cosmological time dilation,  will exceed the lifetime of most observational surveys.  Actually, the situation is not all that bad.  For example, in our 100~\Msun\ helium model the bolometric light curve declines (after peak) at at a rate of about 0.01 mag per day.  At a redshift of $z=10$,  this results in a $\sim0.3$~mag variation over an observer frame year.  That may be within the sensitivity of future surveys with multi-year baselines.   Moreover, we find that the decline is more rapid in the bluer bands  (due to the increasing onset of iron group line blanketing at these wavelengths) such that variations in the restframe U- and B-bands are a factor of $2-3$ larger. Even if the temporal variations are too hard to measure,  the colors of PI~SNe might be distinct enough that one could use them to select candidates for further follow up.

An alternative approach for finding PI~SNe would be to look for the 
brief, but very luminous emission of shock breakout.  For the larger
red-supergiant explosions, the breakout transients are 20 times as
luminous ($L_{\rm peak} \approx 10^{46}~\ergss$) as those of typical
Type~IIP supernovae, and last about four times as long ($\sim
2$~hours).  The emission peaks in the rest-frame far ultraviolet
($\lambda \sim 80-170$~\AA).  Sensitive surveys in the infrared
($2-5~\mu$m) with a high cadence and a wide-field of view
 could, in principle, detect these transients out to high redshifts. 
 Once detected, the subsequent \Nifs\ powered light curve and spectra could be
monitored, with some leisure, for many years to come.


\acknowledgements Support for DK was provided by NASA through Hubble
fellowship grant \#HST-HF-01208.01-A awarded by the Space Telescope
Science Institute, which is operated by the Association of
Universities for Research in Astronomy, Inc., for NASA, under contract
NAS 5-26555.  
AH was supported in part by the US Department of Energy under grant
DE-FG02-87ER40328.
This research has been supported by the DOE SciDAC
Program (DE-FC02-06ER41438).  We are grateful for computer time
provided by ORNL through an INCITE award and by NERSC.


\clearpage

\end{document}